\documentclass[3p,times]{elsarticle}

\usepackage{graphics}
\usepackage{graphicx,subfigure}
\usepackage{amssymb}
\usepackage{amsmath,textcomp,array}
\usepackage{multirow}
\usepackage{color}
\usepackage{ulem}
\usepackage{amssymb}

\usepackage[figuresright]{rotating}

\begin{document}

\begin{frontmatter}




\title{PynPoint Code for Exoplanet Imaging}


\author{Adam Amara}
\author{Sascha P. Quanz}
\author{Joel Akeret}

\address{ETH Zurich, Department of Physics, Wolfgang Pauli Strasse 27, 8093 Zurich, Switzerland}

\begin{abstract}

We announce the public release of PynPoint, a Python package that we have developed for analysing exoplanet data taken with the angular differential imaging observing technique. In particular, PynPoint is designed to model the point spread function of the central star and to subtract its flux contribution to reveal nearby faint companion planets. The current version of the package does this correction by using a principal component analysis method to build a basis set for modelling the point spread function of the observations. We demonstrate the performance of the package by reanalysing publicly available data on the exoplanet $\beta$ Pictoris b, which consists of close to 24,000 individual image frames. We show that PynPoint is able to analyse this typical data in roughly 1.5 minutes on a Mac Pro, when the number of images is reduced by co-adding in sets of 5. The main computational work parallelises well as a result of a reliance on SciPy and NumPy functions. For this calculation the peak memory load is 6Gb, which can be run comfortably on most workstations. A simpler calculation, by co-adding over 50, takes 3 seconds with a peak memory usage of 600 Mb. This can be performed easily on a laptop. In developing the package we have modularised the code so that we will be able to extend functionality in future releases, through the inclusion of more modules, without it affecting the users application programming interface. We distribute the PynPoint package through the central PyPi sever, and the documentation is available online (http://pynpoint.ethz.ch).

\end{abstract}


\end{frontmatter}


\section{Introduction}
\label{sec:intro}

The field of exoplanet research has grown enormously in recent decades, with the current census of confirmed exoplanets and exoplanet candidates exceeding 5,000 objects (see, www.exoplanet.eu \cite{schneider2011}, www.exoplanets.org \cite{wright2011}). The vast majority of the exoplanets known today have been detected through radial velocity (RV) and transit measurements (e.g., \cite{bathala2013,borucki2011,mayor2011,howard2010,cumming2008}). These studies, and in particular the {\it Kepler} space mission, allow us to derive robust statistics for the occurrence rate of planets on close-in orbits as a function of planet size and mass (e.g., \cite{marcy2014,fressin2013,dressing2013,mayor2011}).

In addition to RV and transit searches for exoplanets, a considerable and increasing effort has been devoted to direct imaging searches for exoplanets in recent years. These activities are currently culminating in the commissioning of dedicated high-contrast planet imaging instruments at 8-m class telescopes (SPHERE at the VLT \cite{beuzit2008}, and GPI at Gemini \cite{macintosh2008}). The scientific motivation for direct imaging searches is at least twofold: (1) determining the occurrence rate of exoplanets at larger orbital separations not accessible by RV and transit searches; (2) obtaining spectra and spectral energy distributions (SEDs) of extrasolar planets to constrain their atmospheric parameters. In principle, exoplanet atmospheres can also be studied using transit and secondary eclipse observations (e.g., \cite{seager2010}), but only a small fraction of the exoplanet population is transiting in front of or behind their host star. Examples for directly imaged exoplanets that likely formed from a circumstellar disk around their host star include $\beta$ Pictoris b \cite{lagrange2010}, the 4-planet system around HR8799 \cite{marois2008,marois2010}, HD95086 b \cite{rameau2013b} and GJ504 b \cite{kuzuhara2013}. In addition, in a few cases, there is observational evidence from direct imaging campaigns for very young planets that are embedded in or interact with the gas rich disk of their host stars (e.g. HD100546 \cite{quanz2013}).

The biggest challenge for the direct detection of an exoplanet is the stark brightness contrast between the planet and its host star in combination with the small on-sky separation. Hence, direct planet searches are typically done with ground-based, adaptive optics assisted, very large telescopes at near-infrared (NIR) wavelengths (1 -- 5 $\mu$m), where one aims at the detection of the planet's thermal emission rather than reflected starlight. To enhance the contrast capabilities of the instruments, coronagraphs with different designs have been introduced in the past years, in particular for observations between 3 -- 5 $\mu$m \cite{kenworthy2010,quanz2010,mawet2013}. Furthermore, a big improvement in contrast performance was achieved with the introduction of the angular differential imaging (ADI) technique \cite{marois2006}, which allows for a much more accurate subtraction of the central star's point spread function (PSF).

Together with these technical improvements, it has also been realised that there are significant potential gains from developing better and more sophisticated data analysis methods. A particularly crucial step in the data processing chain is the modelling and subtraction of the star's PSF, as typically the star outshines a planetary companion by orders of magnitude. Hence, only an accurate subtraction of the stellar flux contribution might reveal the existence of a nearby companion. A nice example here was the introduction of the LOCI algorithm \cite{lafreniere2007b} that was developed for the Gemini Deep Planet Search \cite{lafreniere2007} and is still commonly used today. More recently, principal component analysis (PCA) based algorithms were introduced by our group and a group in the US \cite{amaraquanz2012, soummer2012}. As the quality and volume of exoplanet direct imaging data will continue to increase in the coming years, in particular with the advent of the dedicated instruments mentioned above, the analysis methods used to process the data will also have to grow in sophistication. 

In this paper, we give a brief overview of the PynPoint package. We begin with a brief description of the data and processing steps that currently need to be performed before, the data is passed to PynPoint (section \ref{sec:predata}). In section \ref{sec:package}, we give an overview of the PynPoint package; and we describe the main user interfaces in section \ref{sec:API}. We show performance results in section \ref{sec:performance}, and in sections \ref{sec:qa} and \ref{sec:summary} we present a discussion of quality assurance and a summary. 

\section{The Data and Initial Data Reduction Steps}
\label{sec:predata}
Before describing in detail the functionalities of the PynPoint package, it is worth laying out the typical input data that PynPoint uses and the basic data reduction and analysis steps that are carried out before the PSF fitting and subtraction are performed. 

As described above, the standard high-contrast observing technique to image exoplanets is ADI on adaptive optics-assisted NIR cameras. Typically during this process, the pupil of the telescope is fixed throughout the observing sequence, but the stellar field rotates on the detector due to the continuously changing parallactic angle caused by Earths' rotation. Depending on the observing wavelength, the brightness of the star and whether or not a coronagraph is used, the typical exposure time of a single image frame varies between a fraction of a second to several seconds. A typical size of an individual raw image is $1024\times1024$ pixels. However, in some cases smaller detector sub-arrays are read out and stored (e.g., $512\times512$ or 256$\times$256 pixels). A typical observing sequence lasts for up to a few hours, resulting in thousands if not tens of thousands of individual exposures that are stored individually. To be able to correct for bad pixels and sample the PSF more accurately, while properly subtracting background emission, the star is made to cycle through different pre-defined positions on the detector during the observing sequence (`dithering'). The number of exposures taken at a given dither position before the star is moved to the next point depends on the observing wavelengths, the individual exposure time and the observing conditions. While observing at 1 -- 2.5 $\mu$m, one can normally spend up to several minutes at the same dither position. At longer wavelengths (3 -- 5 $\mu$m) and with more strongly varying background (see below), one switches to another dither position every 30 -- 60 seconds. 
 
The raw images undergo several steps before they are processed the PynPoint package. These process include:

\begin{description}

\item[Parallactic angle determination.] Depending on the instrument, not all individual raw images will have a parallactic angle assigned to them. In that case, the parallactic angle for a given raw image has to be computed and stored in the header of the image file. PynPoint will search for a header keyword called {\tt NEW\_PARA} in each individual input image to access the parallactic angle. 

\item[Dark current subtraction.] The dark current (and, where applicable, also the bias level) for a given exposure time and detector readout scheme is subtracted from all raw images. This step can be skipped if in the background subtraction step (see below) two individual images are subtracted from each other, as this will also subtract any dark current signal.

\item[Bad pixel cleaning.] Dead or bad detector pixels, as well as pixels possibly hit by a cosmic ray, have to be identified. These pixels are either masked out and disregarded from further analyses or they are replaced with the mean value of surrounding pixels. 

\item[Flatfielding.] A flat-field correction is applied to the individual images to correct for large scale sensitivity gradients across the detector. In some cases, depending on the required calibration accuracy, this step can be skipped if the detector is reasonably `flat'. 

\item[Background subtraction] Depending on the wavelength, the background emission has to be subtracted from all individual images. This step becomes more important with increasing wavelength and is absolutely crucial for data taken between 3 -- 5 $\mu$m. At these wavelengths, thermal emissions from the sky-background, the telescope and the instrument contribute to the total background flux. Without a nearby bright star, observations at 3 -- 5 $\mu$m are typically limited by the poisson noise of the background flux and temporal changes therein. To subtract the background emission from a given raw image, typically  another image observed close in time but at a different dither position is subtracted. Alternatively, a sequence of several images from a different dither position is mean combined, to reduce the poisson noise in the background, and the resulting average image is removed. It has been suggested, that additional sensitivity can be gained if the LOCI algorithm is used to construct and subtract the background emission in the immediate vicinity of a star for a given input image \cite{galicher2011}.

\item[Image alignment and trimming.] As the star was recored at different dither positions on the detector, the individual images have to be aligned to a common centre. PynPoint requires that all input images are centred on the central star. In order to compute the spatial offset between the different images, either a 2-D Gaussian fit is used to determine the location of the photo-center of the stellar PSF, or the images are cross-correlated. Once the offset is known, the images are scaled up by a factor of at least ten, shifted and then re-scaled back to either their original size or twice that value. In PynPoint, we normally work with images that are up-scaled by a factor of two \cite{amaraquanz2012} as compared to the original pixel scale. This allows us to recover sub-pixel information on the PSF shape. Depending on the data set and the science goals, we typically have image sizes between 1 and 5 arcsec$^2$.   

\end{description}

\begin{figure}[h]
\begin{center}
\includegraphics[width=\linewidth]{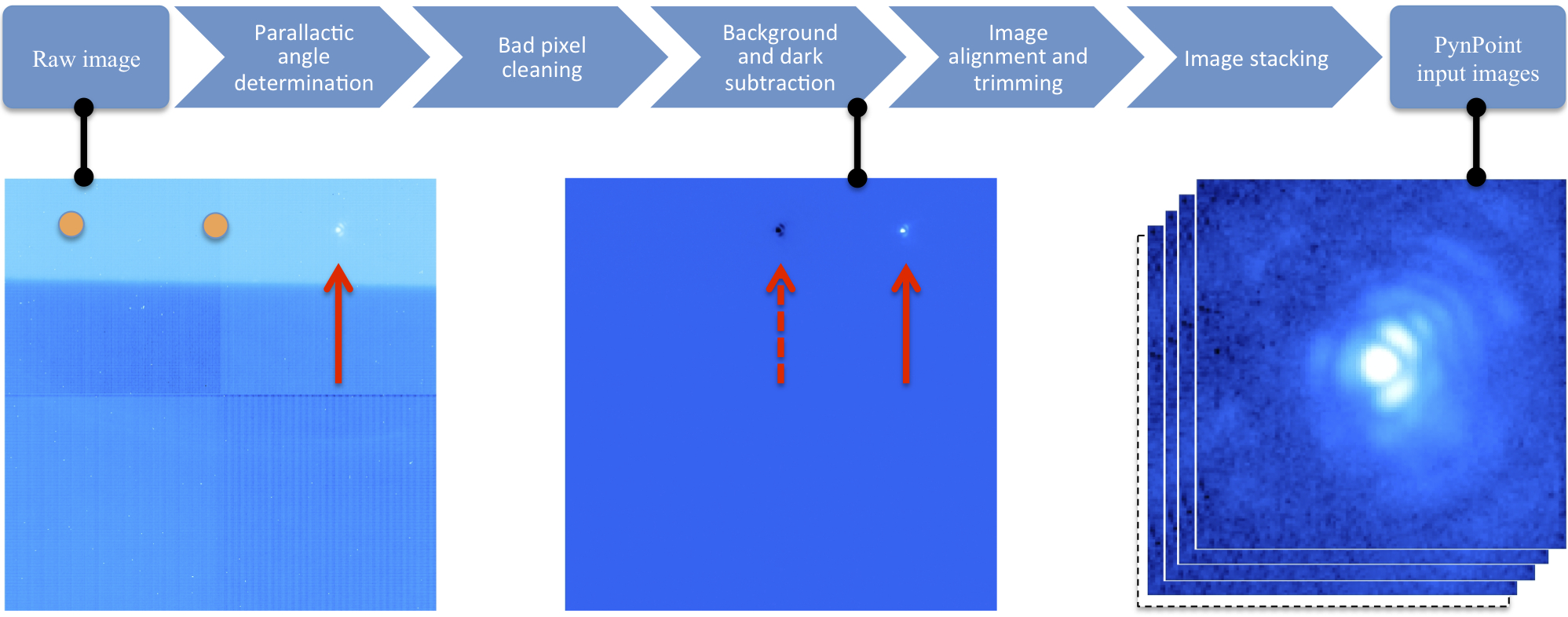}
\end{center}
\caption{Initial data reduction steps for high-contrast imaging data taken with VLT/NACO in the L$'$ filter. In this example, no flat-fielding is applied and the background emission and dark current contribution are eliminated via pairwise subtraction of individual images taken at different dither positions. Also shown is a single raw image (left), an intermediate image after background and dark subtraction step (middle), and a representation of a stack of final input images for PynPoint (right). The raw image is 512x512 pixels in size ($\sim$13$''$x13$''$). The high-background level, as well as the four detector quadrants are clearly visible. Also clearly seen is the bright stripe in the upper quarter of the image, which is related to the design of the APP coronagraph \cite{kenworthy2010}. The current location of the star is indicated by the red arrow, and the other dither positions used in this observing sequence are indicated by the orange circles. In the intermediate image, the detector artefacts are gone and the background appears smooth. In addition to the star (red arrow), a negative imprint of the star is also visible (red, dashed arrow), resulting from the pairwise subtraction procedure. The final input images were created from a mean combination of 10 individual exposures after applying all initial reduction steps. The images are 146x146 pixels in size ($\sim$2$''$x2$''$) and were scaled up by a factor of two with respect to the original image size. The asymmetric PSF of the APP is clearly visible with the diffraction rings being strongly suppressed on the lefthand side of the central star. However, this image also demonstrates that in order to reveal faint companions next to the central star, the stellar PSF needs to be modelled and subtracted as accurately as possible.PynPoint usually processes thousands of such images. }
\label{figure1}
\end{figure}

For different instruments or observing strategies, the exact sequence of the steps listed above may vary slightly and not all steps maybe necessary. In the end, however, PynPoint (v0.2.0) expects a stack of individual .fits files that have all been pre-processed, with the same size in pixels and are aligned properly with the star sitting in the center of the image.

In Figure~\ref{figure1}, we show a typical flow diagram of the initial reduction steps for the data \cite{quanz2010,quanz2011a,quanz2012b,quanz2013}. These data are taken with the NACO camera at the Very Large Telescope of the European Southern Observatory (ESO) in the L$'$ filter (central wavelength $\lambda_{\rm cen}=3.8 \mu$m). In addition to the data reduction steps, Figure~\ref{figure1} also shows a single raw image, an intermediate data product and the final input images for PynPoint for illustrative purposes. In this particular case, the APP coronagraph \cite{kenworthy2010,quanz2010} was used, giving the PSF its asymmetric shape. It has to be emphasised that in each of the recorded images, the PSF of the central star has a slightly different intensity distribution on the detector due to continuously changing observing conditions and related variations in the AO corrections. This variability in the PSF in combination with variations in the uncorrected AO-halo and quasi-static speckles due to non-common path aberrations is the main challenge for PSF-fitting algorithms.

\section{Package Description}
\label{sec:package}
PynPoint is a Python-based package that we have developed for processing high-contrast imaging data. In this first release, we have focused on the steps associated with modelling and subtracting the PSF of the bright central star. However, we have structured the package in a modular way so that extra features can be later added to the package easily. Once new features are developed and tested, we plan to make new releases publicly available. Therefore, an important constraint of the structure is that the released code can be scaled up in a way that maintains backward compatibility. 

Version 0.2.0 of PynPoint contains the modules summarised in Table \ref{tbl:modules}. Among these, the main three classes that will be accessed by the users are {\tt basis, images and residuals}. These classes are used to generate an appropriate basis set for the analysis, treat the data (including modelling the PSF in each frame) and then to subtract the PSF model from each image and average the residuals through the stack of images. 

In order to maintain a clean design, flexible modularity and a clear separation of concerns, we have created a series of modules to manage common features: functionality that is shared amongst multiple classes, such as saving, plotting and animating, has been grouped together in a class hierarchy. In order to facilitate the instantiation and restoration of these classes, a factory module has been developed. Additional generic functionality for manipulating the data has been defined in the {\tt \_Util} helper module.

An important part of the data analysis is to be able to mask regions of the image. The pixels at the centre of the PSF, i.e. the position of the star, are typically saturated. This means that their response to the input flux is far from linear. Hence, the central region is difficult to model accurately, and the information content in this area is minimal to none. Inclusion of these regions can, therefore, degrade the quality of the image analysis. For instance, if the basis set used to model the PSF is orthonormal, inclusion of the central region will affect the basis functions outside the central region. A simple and effective way of dealing with such problems is to mask unwanted regions. Masking operations in PynPoint are handled by the {\tt \_Mask} module.

As well as an interactive mode, discussed further below, we also offer the capability to run PynPoint in a sequence that can be predefined by the user. To manage these operations, we have developed a simple workflow engine (called {\tt workflow}) that can read the user inputs and run the appropriate functions in the right order. 

Detailed instructions of how to download and install the PynPoint package can be found on the package website\footnote{http://pynpoint.ethz.ch}\footnote{http://readthedocs.org/projects/pynpoint} . The simplest method is to use pip\footnote{http://pip-installer.org/}: 

\begin{verbatim}
$ pip install PynPoint-exoplanet --user
\end{verbatim}

This will install the PynPoint package and all of the needed dependencies, which are: NumPy, SciPy, PyFITS, matplotlib and h5py.

\begin{table}[tdp]
\caption{Main PynPoint modules.}
\begin{center}
\begin{tabular}{l|l}
Module & Description \\
\hline
\hline
{\tt Workflow} & Engine for managing a set of PynPoint operations\\
{\tt Basis} & Generates the basis set used to model the PSF\\
{\tt Images} & Manages and prepares the images to be analysed \\
{\tt Residuals} & Calculates and manipulates the residuals between image and model\\
{\tt PynPlot} & Plotting and animation routines\\
{\tt \_BasePynPoint} & Set of base features used in {\tt basis}, {\tt images} and {\tt residuals} \\
{\tt \_Creators} & Factory module\\
{\tt \_Ctx} & For shared resources, such as data, during session \\
{\tt \_Mask} & Provides functions for the masking operations\\
{\tt \_Utils} & Generic functionality for manipulating the data\\
\end{tabular}
\end{center}
\label{tbl:modules}
\end{table}%

\section{Application Programming Interface}
\label{sec:API}

The PynPoint package can be run in different modes that we outline in the sections below. The starting point assumes that there is a directory containing a set of fits files that have already been processed through the pre-PynPoint steps highlighted in Section~\ref{sec:predata}. The .fits files should contain the images centered on the star and each image header should contain a parameter {\tt NEW\_PARA}, which indicates the parallactic angle of this particular image. Example scripts for running PynPoint can be found in the examples folder of the package.

\subsection{Using Interactive mode}

PynPoint can be run from the Python shell using keyword arguments to specify the calculations that should be performed by each module. See Table \ref{tbl:kwargs}
 for a list of the keyword options that can be passed to the classes {\tt images} and {\tt basis}. A typical session can be executed as follows:

\begin{verbatim}
>>> import PynPoint
>>> images = PynPoint.images.create_wdir(dir1,cent_size=0.05,stackave=5)
>>> basis = PynPoint.basis.create_wdir(dir2,cent_size=0.05,stackave=5)
>>> res = PynPoint.residuals.create_winstances(images, basis)
\end{verbatim}

It is possible to create a {\tt basis} instance with some of the keyword options different from those used to produce {\tt images}. In general, this is likely not to be a sound idea and should be avoided. Therefore, users should take some caution to understand what they are trying to test if they use different options. In the example above, an instance of the {\tt image} class is created by reading in data from the directory (`dir1'). This option reads in all the files in this directory that end in .fits and stores all the images in the array {\tt im\_arr}, which will have dimensions $N_x \times N_y \times N_{files}$, where $N_x$ and $N_y$ are the number of pixels in the x and y directions of each image and $N_{files}$ is the number of files in the directory. We find that reading in a large number ($\sim$10,000) of small fits files can be very time consuming. Hence, instead of reading these files each time a particular directory is targeted, PynPoint converts the data into an HDF5 file\footnote{http://hdfgroup.org/HDF5}. This file is stored inside the original target directory and is given the same name as the directory with the hdf5 extension. In future calls, if the data in this directory is needed, PynPoint first checks to see if such a HDF5 file exists in the directory. If it does, then the data is loaded directly from the HDF5 file instead of the individual files. 

The images stored (in the attribute {\tt im\_arr}) are: (i) normalised so that the sum of the pixel values is one; (ii) resized/re-centered (optional); and (iii) masked. The operations are managed by the function {\tt prep\_data} in the {\tt \_Util} module. All of the information needed to reverse these steps and to recover the original input data is contained in the {\tt images} instance. 

The {\tt basis} class is designed to generate and manage the basis set that will be used to model the PSF. In the current version, this is done by constructing a set of principal components using singular value decomposition (SVD). For details on the performance of this basis set, see  \cite{amaraquanz2012,2014ApJ...780...17M}. However, we have structured the code in such a way that different basis set options can be added to PynPoint in the future if desired. The basis set information is stored in the {\tt pdf\_basis} attribute of the basis instance as a NumPy array.

The {\tt residuals} class is designed to take data from {\tt images} and {\tt basis} instances and to perform the steps that involve: (i) modelling the PSF; (ii) subtracting the PSF from each frame; (iii) rotating the ADI images to a common sky orientation; (iv) averaging over the stack; and (v) smoothing the images. The most critical input for these calculations is the number of basis functions that should be used to model the PSF. Since the image manipulation steps can be time consuming, we have built in a feature that will store the PSF model information when a given number of basis coefficients is used. This means that any subsequent residual method calls that use the same number of basis functions can by-pass the computationally intensive step and use the previously calculated result directly. However, if the user does a computation that changes the number of basis coefficients, the function will perform a new calculation to model the PSF.

The {\tt  basis, images, residuals} and {\tt workflow} classes offer a save feature, which stores all the data of the instances. This allows us to then reconstitute an object using a restore feature. For instance, in the following example {\tt images} and {\tt images2} will contain the same information:

\begin{verbatim}
>>> images.save(filename)
>>> images2 = PynPoint.images.create_restore(filename)
\end{verbatim}

\begin{table}[tdp]
   \centering
   \begin{tabular}{l|l|l} 

      Option      & Default & Description \\
\hline
\hline
      recent        &  False     &  If True, the images will be re-centered. \\
      resize       &  False & If True, the final images will be increased by a factor set by F\_final. \\
      cent\_remove  & True  & If True, the central region will be masked (size set by cent\_size). \\
      para\_sort & True    & If True, the images will be sorted so that the parallax angles increase through the stack. \\
      F\_final & 2 & Factor increase in resolution of final images (resize must be True).\\
      cent\_size & 0.05 & Radius of the central mask as fraction of the full image size. \\
      edge\_size & 1.0 & Diameter of the outer mask in fraction of the image size. \\
                         &  &(1.0 corresponds to circle that touches the edge of the image).\\
      stackave & None & If set to an integer N, then the stack will be reduced by averaging over adjacent N images.\\
      ran\_sub &None& If set to an integer M, then a random subset of this size will be analysed. \\
      
   \end{tabular}
   \caption{Main keyword options used by {\tt images} and {basis} along with their default values.}
   \label{tbl:kwargs}
\end{table}

\begin{figure}[t]
\begin{center}
\includegraphics[width=2.1in]{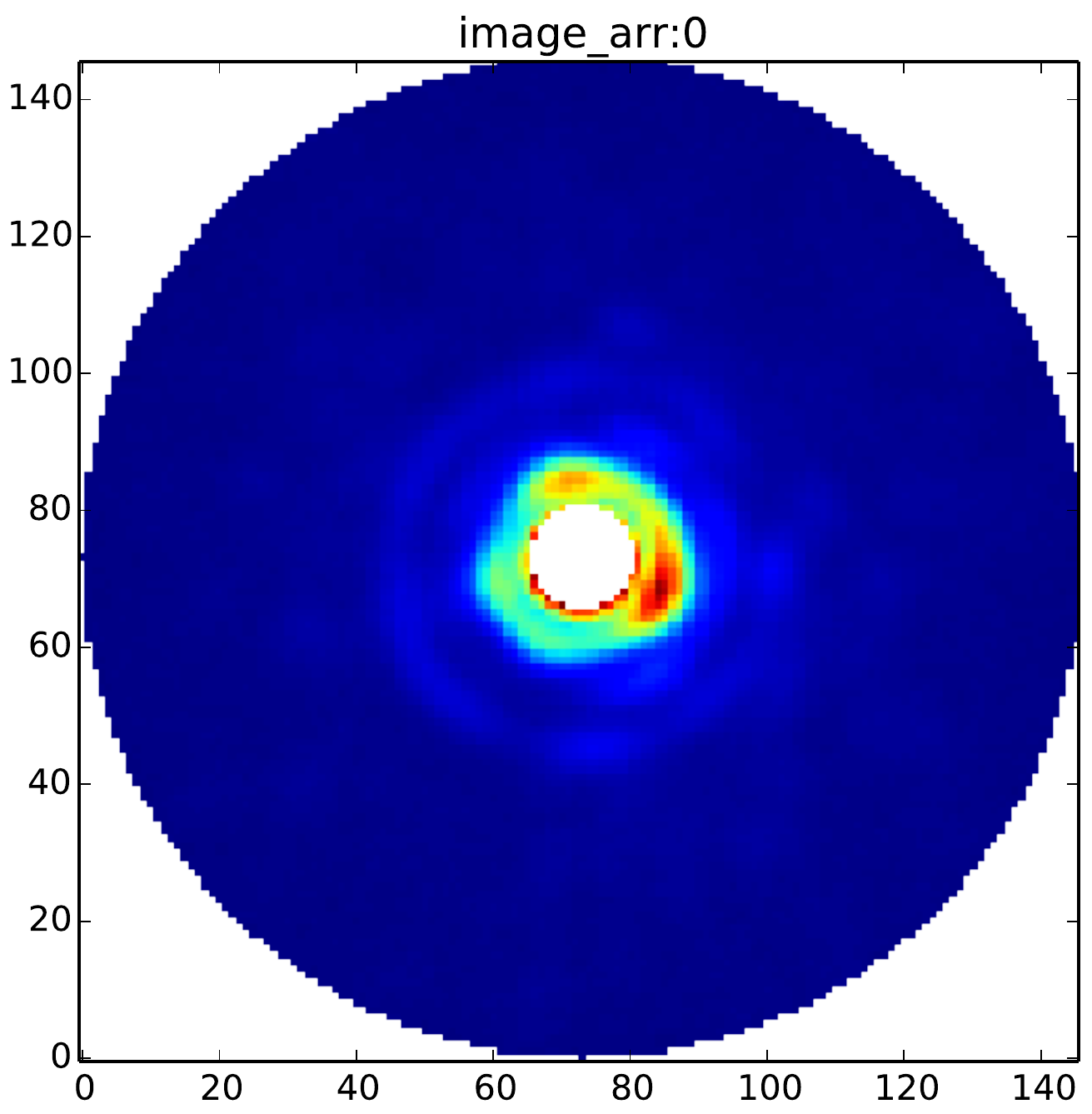}
\includegraphics[width=2.1in]{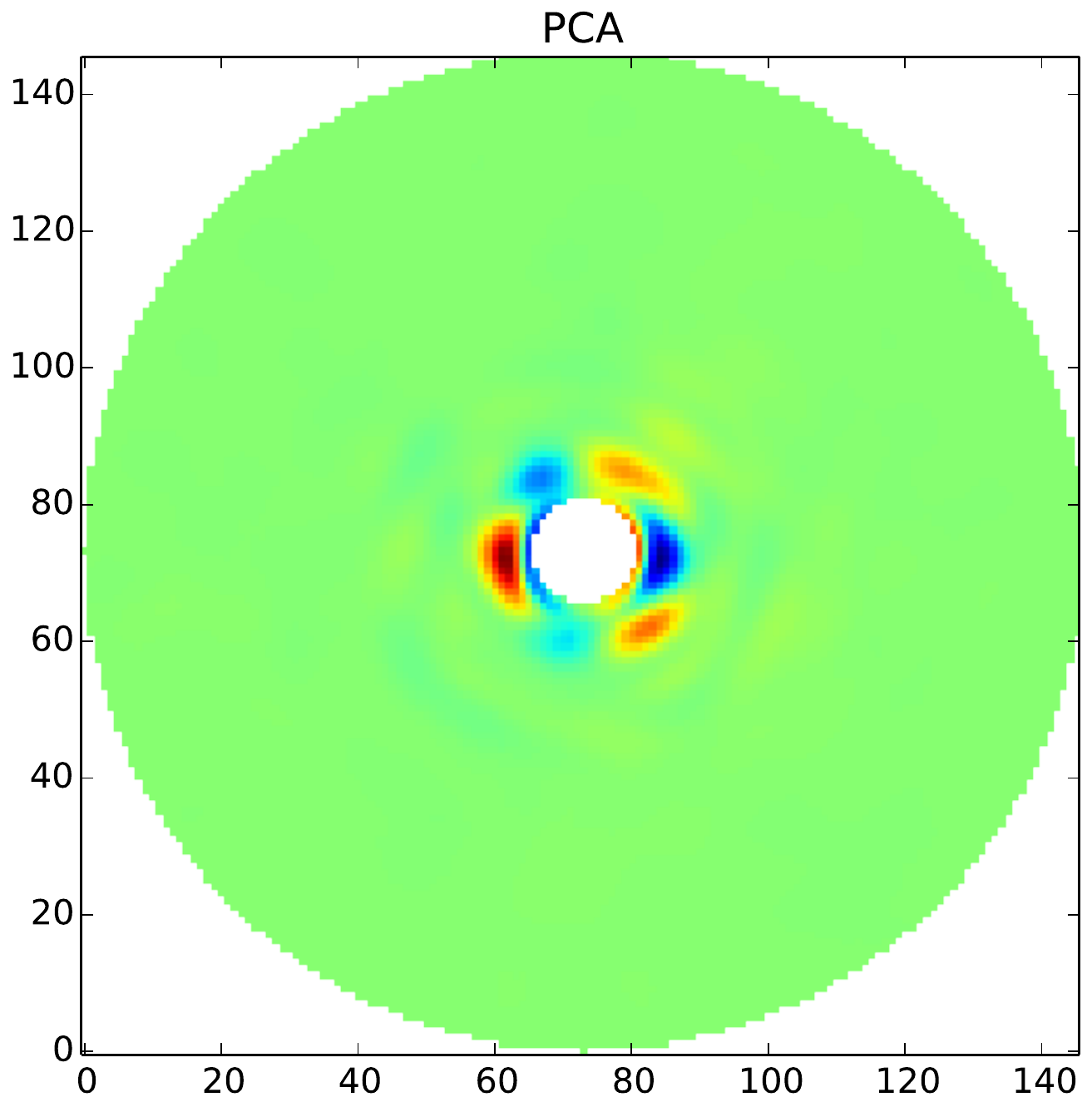}
\includegraphics[width=2.1in]{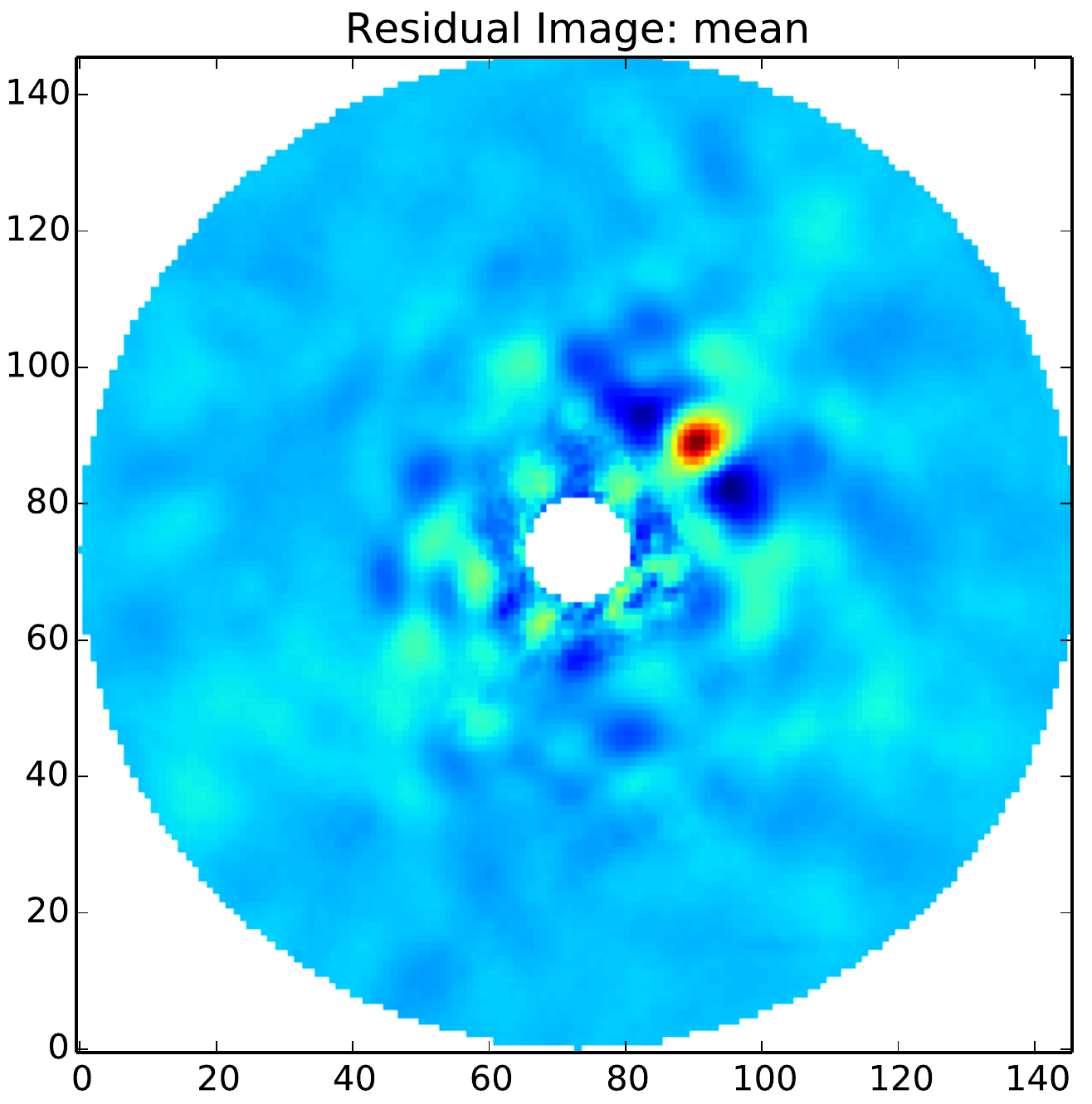}

\end{center}
\caption{Results plotted using the {\tt pynplot} module. The panel on the left shows one of the 2D images stored in {\tt images}. A mask has been applied (centre and corners), and the images have been normalised and the mean image through the stack has been subtracted. The central panel shows the second principal component stored in {\tt basis}. Finally, right panel shows the final image after correction using the first 20 principal components. The red object to the top-right of centre is the exoplanet $\beta$-pic b. In all panels, the colour scale is in arbitrary units. }
\label{fig:pynplot_eg}
\end{figure}

The data produced and stored in various instances can be accessed directly and plotted using functionality of the matplotlib library. We have also included a plotting module that allows users to easily produce some of the most common figures in a standard way.  As an illustration, the panels of Figure \ref{fig:pynplot_eg} can be produced as follows: 
\begin{verbatim}
>>> from PynPoint import PynPlot
>>> PynPlot.plt_im_arr(images, 0)
>>> PynPlot.plt_psf_basis(basis,1)
>>> PynPlot.plt_res(res,20)
\end{verbatim}

The plotting function {\tt plt\_res} accepts a number of keywords such as `smooth', which can be used to smooth the final images through a convolution with a Gaussian of a specified width, and `extra\_rot', which can be used to rotate the images so that (for instance) North points up. The keyword options available to the users are described in our online documentation. The {\tt PynPlot} module can also be used to output a fits file containing the data plotted. This is useful in case users want to examine the images with other packages such as DS9\footnote{http://ds9.si.edu}. 

{\tt PynPlot} also includes functions for showing animations through the image stack using the {\tt anim\_im\_arr} function.

\subsection{Using the workflow engine}

\begin{figure}[t]
\begin{center}
\includegraphics[width=5in]{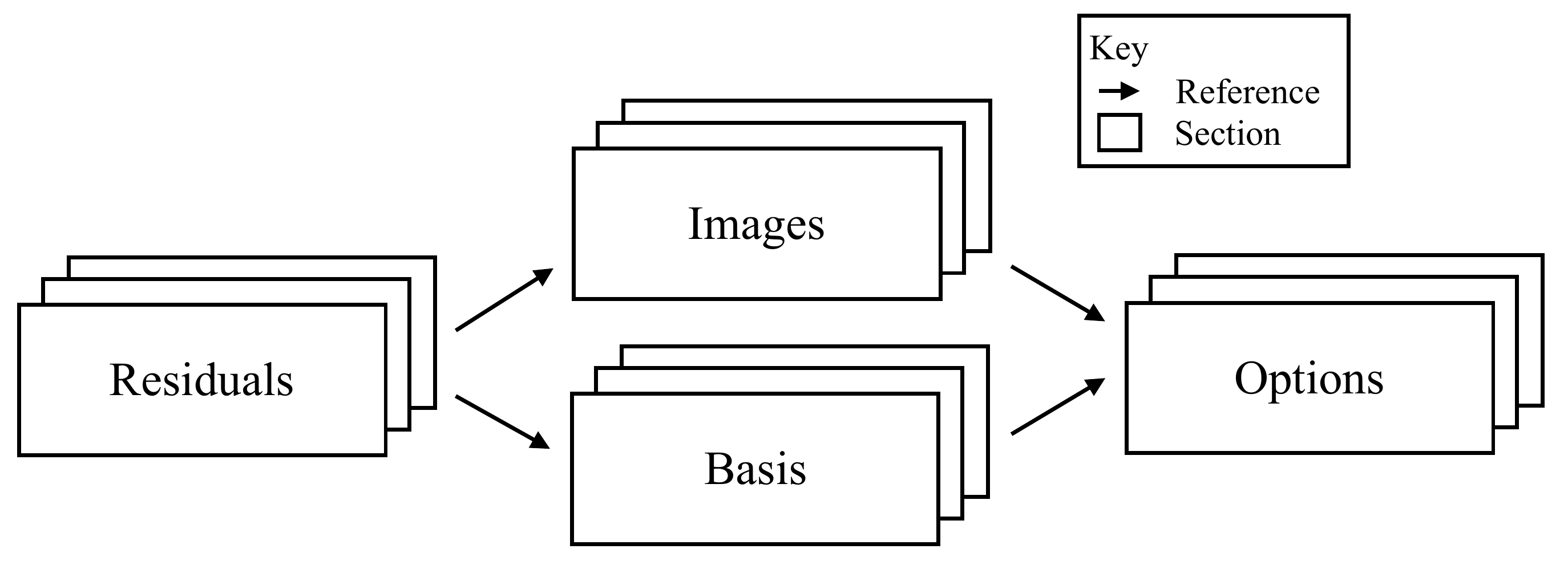}
\end{center}
\caption{The flexible design of the workflow configuration files allows for customizing the behaviour of the {\tt images}, {\tt basis} and {\tt residuals} instances. We show that a given config file can consist of a large number of sections. Several {\tt residual} instances can be generated, where each one takes inputs from an {\tt images} and a {\tt basis} instance. The sections associated with each {\tt images} and {\tt basis} calculation themselves take their inputs from sections that define the user specified options.}
\label{fig:config}
\end{figure}

Given that users may wish to process several data sets with a number of options, relying exclusively on an interactive mode may not always be ideal. For this reason, we have developed a simple workflow engine for managing PynPoint runs. The {\tt workflow} class takes in a configuration file, where the users can specify all the operations that should be run along with keyword options. Figure \ref{fig:config} shows the design of a PynPoint configuration file. It allows users to define a set of {\tt images} and {\tt basis} configurations, each may have a custom {\tt options}  definition,  and a set of {\tt residuals} configurations referring to other definitions. An example of such a config file is shown in \ref{app:config} and further examples can be found in the examples directory of the package. 

Dealing with data volumes can be one of the most difficult challenges in analysing a series of target stars. To gain speed, the simplest approach is to load all the data into memory. However, there is a risk that the data volumes can become too large, compared to the available RAM. In this situation, computer systems may begin to do memory swapping to the disk, which causes a catastrophic slowing of calculations. 

To allow for variability in the data generated by the workflow, the engine has a flexible design. A key feature of this design is to reduce the coupling between the executed modules. Therefore we introduced the concept of a context object, which adds an abstraction layer to the way the data is being accessed and stored. This level of separation will allow us to easily add more sophisticated features to manage large runs, such as caching, without having to interfere with the main PynPoint functionality. This will yield better backward compatibility, as well as robustness of the implementation.

Similar steps to the ones executed in the interactive example above can be performed using the config file in \ref{app:config} as follows,

\begin{verbatim}
>>> import PynPoint
>>> ws = PynPoint.run(configfile, force_replace=True)
>>> res = ws.get(`residuals_module3')
\end{verbatim}

These operations will lead to a {\tt residual} instance that can then be manipulated and plotted in the usual way. In performing the calculations, {\tt workflow} creates a workspace (ws), where all the important information about the run is stored. Setting the {\tt force\_replace} option to True means that previous versions of the workspace with the same name, if they exist, will be replaced. The default setting ({\tt force\_replace = False}) will raise an error if the {\tt workdir} stated in the config file already exists. Knowing the workspace directory ({\tt workdir}  = dir\_ws), it is possible to restore and recover the intermediate information:

\begin{verbatim}

>>> import PynPoint
>>> ws = PynPoint.restore(dir_ws)
>>> images = ws.get(`images_module1')
>>> basis = ws.get(`basis_module2')
>>> res = ws.get(`residuals_module3')
\end{verbatim}

Here, the instances of {\tt images}, {\tt basis} and {\tt res} have been recovered using the {\tt get} method. The available instances in {\tt ws} can be printed

\begin{verbatim}
>>> ws.get_available()
\end{verbatim}

One further advantage of working with the workflow manager is that communicating the data analysis steps to the rest of the science community becomes simple. Authors need only to include a copy of their PynPoint configuration file and the steps taken will be transparent. If the data is also made available, then the results can be easily reproduced.

\subsection{Command line execution of the workflow engine}

It is often useful to be able to launch jobs from the command line. This is usually the case when dealing with a queueing system on a computing cluster. To simplify this process, we have also provided a command line interface that can run the workflow engine in a terminal by passing the name of the config file to be used. 

\begin{verbatim}
$ PynPoint <configfile> True
\end{verbatim}

Including the True at the end has the affect of setting the {\tt force\_replace} option to True. Once completed, the results can be loaded in the same way as previously introduced.

\section{Performance}
\label{sec:performance}

In the following section we report on the performance of the PynPoint package. To do this, we first define our benchmark data set. For continuity and consistency, we have decided to process the same data that we had used to report the science performance of our PCA method \cite{amaraquanz2012}. These data were initially used to confirm the massive planet orbiting $\beta$ Pictoris \cite{lagrange2010}. The dataset was taken with the VLT NACO instrument on 26 December 2009. The observations were made in the L$^\prime$ filter in ADI mode, and 80 data cubes, each with 300 individual exposures, were saved. The field rotation over the observations was roughly $44^\circ$. The initial data reduction steps, corresponding to the steps described in Section \ref{sec:predata}, can be found in \cite{quanz2010}. As part of the re-centering step, the resolution of the images was doubled. Therefore, the final images used in PynPoint ($146\times146$ pixels) correspond to the central ($73\times73$) detector pixels around the central star. After the removal of bad images, 23,839 images remain corresponding to a total data volume (in .fits files) of 3.2 Gbs.

\begin{figure}
\begin{center}
\includegraphics[width=5in]{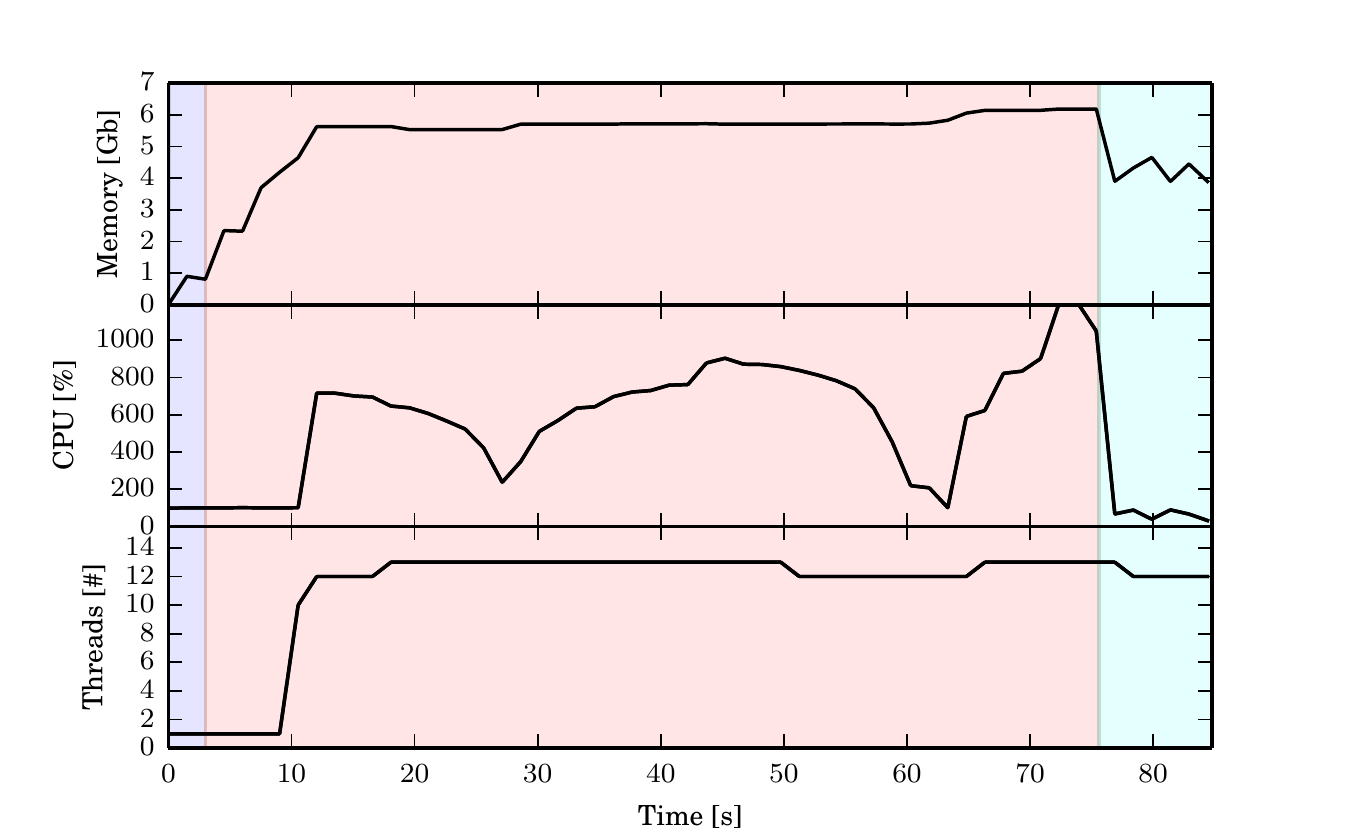}
\end{center}
\caption{Performance profile for our benchmark calculation. The top-panel shows the memory used in Gb; the middle panel shows the percentage cpu with 100\% corresponding one core, and the bottom panel shows the number of threads. The different coloured regions show the various phases of the workflow. From left to right, these calculations are for: {\tt images} (blue), {\tt basis} (red), {\tt residuals} (green - not visible) and the phase where the data is being saved to the disk (cyan). The entire calculation is completed in 1.5 minutes.}
\label{fig:performance_temp}

\end{figure}

We ran a bench mark calculation using the configuration and options shown in \ref{app:config}. These were run on a Mac Pro with an Intel Xeon E5 Ivy Bridge 12-core (2.7 GHz) and 64GB (4x16GB) of 1866MHz DDR3 memory. The results are shown in Figure \ref{fig:performance_temp}. The blue region on the left shows the time spent on calculations of the {\tt images} module and the red region shows the time for the {\tt basis} module. Following this, there is a small (not visible) period while the {\tt residual} instance is being set up. The final phase, in cyan, tidies up the results and outputs the results to disk. We see that the calculations presented here take roughly 1.5 minutes of wall time to complete using {\tt stack\_ave = 5} and that the most computationally intensive parts of the calculations are inside the basis phase, where the principal components are being calculated. The principal component computation relies on the SciPy function {\tt linalg.svd} that, as can be seen by the figure, is automatically parallelised through Intel MKL\footnote{https://software.intel.com/en-us/intel-mkl} and is able to use multiple cores.  

To achieve this run time of 1.5 minutes, we have relied on the fact that we load all of the relevant data into Python and we try to maximise our use of vector and matrix operations. While being fast, the downside of this approach is that the memory demands can be high. In the run we show the memory peak is around 6 Gb. On workstations this should not be a problem, even for large datasets, but smaller machines - such as laptops - will need to run less demanding jobs (for example by increasing the number of frames in a stack). For comparison, a similar run with {\tt stack\_ave = 50}, which should offer reasonable results \cite{2014ApJ...780...17M}, takes a wall time of 3 second and a peak memory load of 600 Mb. On the other hand, it is also possible to perform the calculations with no stacking. For this example, this was done within 40 minutes with a peak memory of roughly 30Gb, which can be accommodated on powerful workstations. In future versions of PynPoint, we will aim to reduce the memory consumption by improving the memory management, which should make large jobs even easier to run.

For the residual calculations, we load the results that are calculated from the workflow run (using the {\tt ws.get} command). Calculating the final residual image for a given number of basis coefficients takes roughly 20 seconds the first time. We have implemented a rudimentary caching scheme so that if the user wishes to calculate other properties from residual files with the same number of basis coefficients, then the calculations are much faster (i.e. $< 0.01$ seconds for a repeat calculation). 

\section{Quality Assurance}
\label{sec:qa}

In order to assure a high level of code quality, we have adopted a test driven development methodology, where unit tests are written early in the code development process, ideally before the main code has been written. This helps us write cleaner and well-tested software and ensures that newly developed features do not affect previously written code. The PynPoint unit tests have been developed with the py.test testing suite\footnote{ http://pytest.org}. Much effort has been put into maintaining a high level of test coverage with a focus on ensuring the numerical correctness of the output. Additionally, we tried to test the functionality and interface exposed to the user by taking a best effort approach and ensure that functions test the inputs they are given through assert commands. Furthermore we  have also adopted a continuous integration approach. Allowing for automated execution of unit tests and reporting after check in of new functionality ensuring a stable state of PynPoint in our code versioning system at any time. We do this by using the publicly available software Jenkins\footnote{http://jenkins-ci.org}.

\section{Summary}
\label{sec:summary}

We have given a description and brief performance illustration of our Python package PynPoint. This package has been written to analyse imaging data for the direct searches of exoplanets. The current version of the code relies on a principal component analysis method to model the point spread function of the central star. This then allows us to subtract the flux of the star and to search the images for faint companions. We have shown that for typical calculations and data volumes, the PynPoint steps can be calculated on the time scale of roughly 1.5 minutes. 

We are distributing the code through the central PyPi server\footnote{https://pypi.python.org/pypi/PynPoint}, which makes our package easy to access and install. We will also maintain documentation for the package at read-the-docs (see http://pynpoint.ethz.ch), which we also use as the project website. As well as the static content delivered through the website, we have also created a wiki\footnote{https://wiki.phys.ethz.ch/PynPoint} for more dynamic content. Users are encouraged to join the PynPoint mailing list (pynpoint@lists.phys.ethz.ch), which can be done by following the instructions on the PynPoint website. Users keen to contribute to the project should feel free to contact us. Contact details can be found on the PynPoint websites.

\section*{Acknowledgements}

We would like to thank Alexander Refregier, Lukas Gamper and Laurenz Gamper for their support through the ETH Cosmology Software Lab. We also thank Michael R. Meyer, Henning Avenhaus, Conor Francois and Maddalena Reggiani for useful discussions in the development and testing of the current release of PynPoint. 
 
\newpage

\appendix

\section{Config file example}
\label{app:config}

\begin{verbatim}
[workspace]
workdir = ../data/workspace_betapic_stk5/
datadir = ../data/

[module1]
mod_type = images
input = Data_beta_pic_L_Band/
intype = dir
options = options1

[module2]
mod_type = basis
input = Data_beta_pic_L_Band/
intype = dir
options = options1

[module3]
mod_type = residuals
intype = instances
images_input = module1
basis_input = module2

[options1]
cent_remove = True
cent_size = 0.05
edge_size = 1.0
resize = False
F_final = 2  # Not used if resize is not True
recent = False
para_sort = True
stackave = 5

\end{verbatim}

\newpage

\bibliographystyle{elsarticle-num}
\bibliography{mybib}







\end{document}